\documentclass[conference]{IEEEtran}
\IEEEoverridecommandlockouts

\usepackage{cite}
\usepackage{amsmath,amssymb,amsfonts}
\usepackage{algorithm}
\usepackage{graphicx}
\usepackage{algorithmicx}
\usepackage{algpseudocode}
\usepackage{subcaption}
\usepackage{textcomp}
\usepackage{xcolor}

\def\BibTeX{{\rm B\kern-.05em{\sc i\kern-.025em b}\kern-.08em
    T\kern-.1667em\lower.7ex\hbox{E}\kern-.125emX}}
    \usepackage{fancyhdr}
\usepackage{bm}        

\begin{document}
 \title{A Blockchain-Enabled Framework of  UAV Coordination for Post-Disaster Networks}
 \author{\IEEEauthorblockN{Sana Hafeez, Runze Cheng, Lina Mohjazi, Muhammad Ali Imran and Yao Sun\\}
\IEEEauthorblockA{\textit{James Watt School of Engineering, University of Glasgow, Glasgow, United Kingdom} \\
     Email: {\{s.hafeez.1, r.cheng.2\}@research.gla.ac.uk},\\ {\{Lina.Mohjazi,   Muhammad.Imran, Yao.Sun\}@glasgow.ac.uk}}}
\maketitle
\begin{abstract}
Emergency communication is critical but challenging after natural disasters when ground infrastructure is devastated. Unmanned aerial vehicles (UAVs) offer enormous potential for agile relief coordination in these scenarios. However, effectively leveraging UAV fleets poses additional challenges around security, privacy, and efficient collaboration across response agencies. This paper presents a robust blockchain-enabled framework to address these challenges by integrating a consortium blockchain model, smart contracts, and cryptographic techniques to securely coordinate UAV fleets for disaster response. Specifically, we make two key contributions: a consortium blockchain architecture for secure and private multi-agency coordination; and an optimized consensus protocol balancing efficiency and fault tolerance using a delegated proof of stake practical byzantine fault tolerance (DPoS-PBFT). 
Comprehensive simulations showcase the framework's ability to enhance transparency, automation, scalability, and cyber-attack resilience for UAV coordination in post-disaster networks. 

\end{abstract}

 \section{Introduction}
Unmanned aerial vehicles (UAVs) offer unique advantages for rapidly restoring communication and coordinating emergency response when ground networks are damaged after disasters \cite{b1}. With agile mobility and aerial positioning, UAV networks can provide ad hoc connectivity among affected areas to assess damage, deliver critical supplies, and assist search-and-rescue efforts \cite{b2}. Deploying UAV fleets addresses situational challenges like blocked access and downed infrastructure that hinder conventional emergency relief \cite{b3}. Leveraging these capabilities requires addressing coordination limitations in uniting heterogeneous, decentralized UAV platforms across agencies. Integrating blockchain architecture with UAV networks shows promise in addressing these limitations for decentralized disaster response \cite{b4}.

Recent studies have proposed integrating blockchain with UAVs to enable decentralized emergency coordination \cite{b5}. While promising, existing blockchain-UAV solutions lack comprehensive security assessments against threats targeting disaster relief systems \cite{b6}. Additional open challenges include optimizing consensus protocols, preserving privacy, designing adaptive smart contracts, and evaluating resilience against failures and attacks \cite{b7}.
Timely disaster response also requires coordinating tasks across affected areas. However, challenges like damaged infrastructure, security vulnerabilities, centralized visibility, and inter-agency collaboration issues persist with conventional ground-based approaches. UAVs offer more agile coordination but have limitations like short flight times, reliance on compromised networking, and inability to securely share data across platforms \cite{b8}.
Integrating blockchain with UAVs for disaster response introduces key research gaps. These include developing optimized consensus protocols for resource-constrained UAVs, enabling interoperability across heterogeneous fleets, evaluating security against threats, and designing context-aware smart contracts \cite{b9}.

This research is motivated by addressing the limitations above through optimized blockchain architecture and protocols tailored for decentralized coordination of heterogeneous UAV fleets amidst disasters. 
Our research makes the following key contributions:
\begin{itemize}
    \item We introduce a blockchain architecture that enables decentralized coordination across different UAV fleets. This system is designed with a focus on preserving privacy and access control, which is crucial for efficient and secure operations in disaster response scenarios.
    \item We develop an optimized consensus protocol that synergizes DPoS with PBFT. This protocol is specifically tailored for UAV networks, aiming to achieve lightweight processing, high throughput, low latency, and robust fault tolerance.
    \item We analyze how overcoming the limitations of current UAV network coordination technologies can unlock the full potential of decentralized, intelligent UAVs for disaster response. This motivates further research and development in blockchain techniques to enable resilient, collaborative, and autonomous UAV-based operations in emergency scenarios.
\end{itemize}
\begin{figure}[ht!]
\centering
\includegraphics[width=0.50\textwidth]{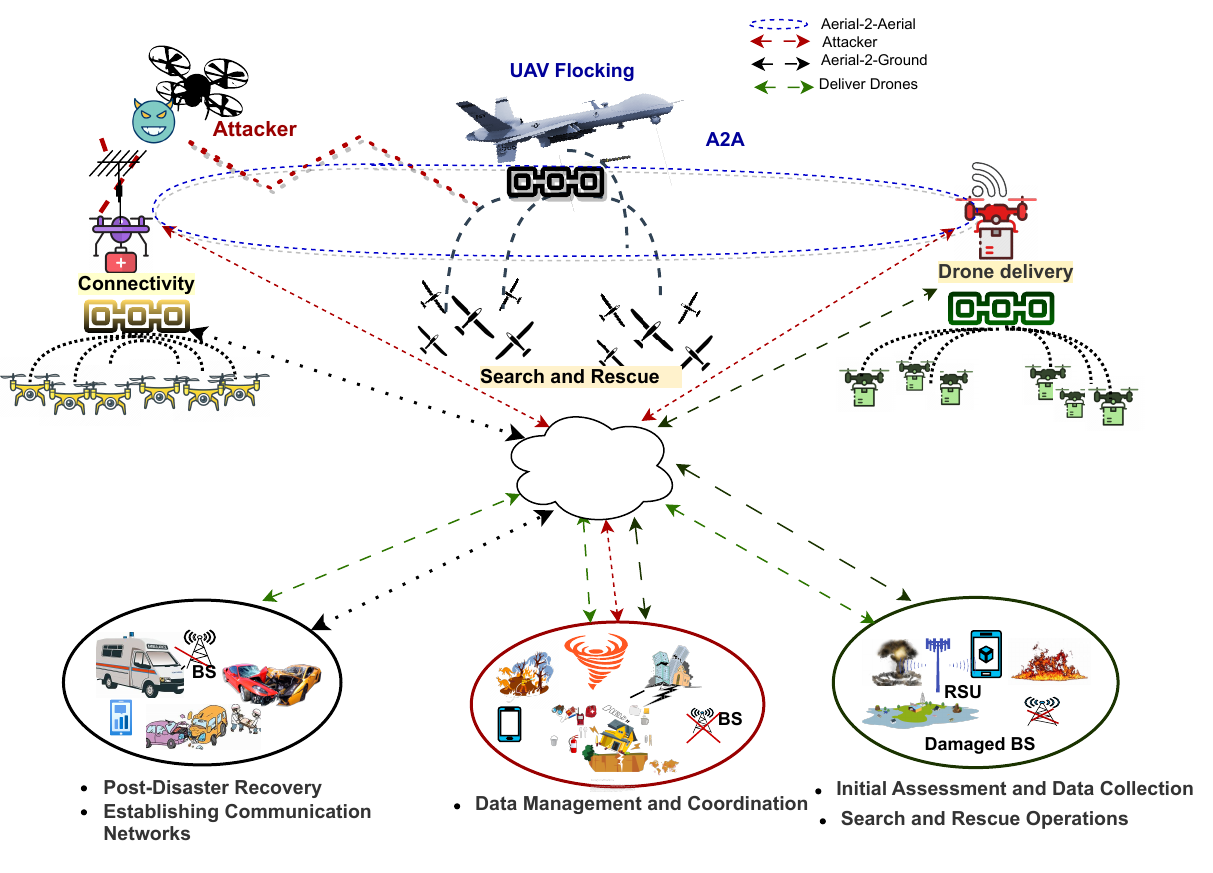}
\caption{Architecture for Blockchain-Enabled UAV Coordination in Disaster Response.}
\label{fig:architecture}
\end{figure}
The structure of the paper is laid out as follows: Section \ref{sec:model} describes the proposed system architecture and design. Section \ref{sec:hybrid_consensus} then introduces our consensus protocol, combining aspects of multiple popular blockchain consensus mechanisms. Section \ref{sec:results} delves into the simulation setup, results, and analysis of the performance of our proposed approach. Finally, Section \ref{sec:conclusion} summarizes the contributions of the paper and suggests potential avenues for future work.
\section{System Architecture and Models}
\label{sec:model}

We consider a disaster response scenario with a heterogeneous fleet of UAVs deployed for search, connectivity, delivery, and assessment operations. The system architecture comprises aerial vehicles, ground support infrastructure, and a central command unit for coordination, as shown in Fig. \ref{fig:architecture}. 
The air segment has 200 UAVs clustered by mission type: 50 provide LTE connectivity, 100 conduct supply delivery, 25 perform search and rescue tasks, and 25 assess the damage. Intra-cluster flocking algorithms enable anti-collision and coordinated mobility within mission groups. Inter-cluster protocols share situational awareness and dynamic task allocation across groups.
Ground infrastructure includes roadside units (RSUs) that offer connectivity access points to first responders. The command center monitors (C2) overall progress through networked links to air and ground assets, directing planning, and adapting swarm activities based on evolving needs. Efficient aerial response necessitates optimizing performance across several axes, including communications, control algorithms, reliability, security, and latency. Mathematical models guide analysis in these aspects. Having described the overall system architecture, next we delve deeper into defining the communication model for data exchange between UAV clusters and ground infrastructure.
\subsection{Communication Model}
\label{sec:comm_model}

The communication model defines the key parameters for reliable data exchange among the UAV nodes. Building on this communication framework, we next examine the consensus protocol for securely coordinating actions and sharing information between UAVs. The signal-to-noise ratio (SNR) determines the channel capacity for UAV-to-UAV links based on transmit power, antenna gains, separation distance and noise density as per equation \ref{eq:snr}.
\begin{equation}
\label{eq:snr}
\mathrm{SNR}{ij} = \frac{P{t_{ij}} G_{t_i} G_{r_j} \lambda^2}{(4\pi d_{ij})^2 N_0},
\end{equation}
Where \( P_{t_{ij}} \) is the transmit power from UAV \( i \) to UAV \( j \), \( G_{t_i} \) and \( G_{r_j} \) represent the transmit and receive directional antenna gains, respectively, \( \lambda \) is the carrier wavelength, \( d_{ij} \) is the inter-UAV distance, and \( N_0 \) is the noise power spectral density. The maximum achievable data rate \( C_{ij} \) between nodes is calculated using the Shannon-Hartley theorem relation in equation \ref{eq:shannon}, which depends on the SNR and allocated bandwidth $B_{ij}$.

\begin{equation}
\label{eq:shannon}
C_{ij} = B_{ij} \log_2(1 + \mathit{SNR}_{ij}).
\end{equation}

A dynamic bandwidth allocation scheme adapts the capacity across inter-UAV links based on changing channel conditions and data demands. Maintaining an up-to-date global state is vital for coordination. Cluster heads act as leaders in disseminating commands among follower UAVs using a fault-tolerant consensus algorithm resistant to disruptions. The permissioned blockchain framework facilitates secure data replication using access controls, smart contracts, and cryptographic signatures.

\subsection{Mobility Model}
To realistically simulate UAV movement patterns in response scenarios, reflecting their operational constraints and mission objectives.
A modified Random Waypoint Model (RWM) is used, incorporating velocity vectors and acceleration constraints. The position of UAV \(i\) at time \(t\) is given by
\begin{equation}
\mathbf{P}_i(t) = \mathbf{P}_i(t-1) + \mathbf{V}_i(t) \Delta t + \frac{1}{2} \mathbf{A}_i(t) \Delta t^2.
\end{equation}
where \( \mathbf{V}_i(t) \) and \( \mathbf{A}_i(t) \) are the velocity and acceleration of UAV \(i\) at time \(t\), and \( \Delta t \) is the time step. Building on this mobility model, we next conduct an analysis of the key latency components that impact real-time communication efficiency in UAV networks.
\subsection{Analysis of Latency Components in UAV Networks}

The overall communication latency in UAV networks comprises of four key components:

\begin{enumerate}
\item Processing Latency ($L\text{proc}$): The time taken by the UAV hardware to perform computational tasks like signal processing, encryption etc. It depends on the capability of the on-board processors.
\begin{equation}
    L_{\text{proc}} = \text{Time taken for computational tasks.}
\end{equation}
\item Queuing Latency ($L\text{queue}$): The time data packets spend waiting in queues to be processed or transmitted. Influenced by network traffic, UAV data handling capacity, and packet priorities.
\begin{equation}
    L_{\text{queue}} = \frac{\text{Average queue length}}{\text{Service rate}}.
\end{equation}

\item Transmission Latency ($L_\text{trans}$): Time taken to transmit data over communication channels. Calculated using the Shannon-Hartley theorem relating channel capacity, bandwidth and SNR.
represents the baseline noise floor.
\begin{equation}
    L_{\text{trans}} = \frac{1}{C} = \frac{1}{B \log_2(1 + \text{SNR})}.
\end{equation}

\item Propagation Latency ($L_\text{prop}$): Time for signals to propagate from sender to receiver UAV based on their separation distance. Propagation latency, often a smaller component in aerial networks, is the time taken for a signal to travel from the sender to the receiver. It is calculated using the distance between the UAVs and the speed of light:
\begin{equation}
    L_{\text{prop}} = \frac{d_{ij}}{c}.
\end{equation}
\end{enumerate}

The total latency is the sum of these individual components

\begin{equation}
L_\text{total} = L_\text{proc} + 
L_\text{queue} + L_\text{trans} + \
L_\text{prop}
\end{equation}

In particular, transmission and propagation latencies significantly impact the real-time communication efficiency in UAV networks for disaster response scenarios. We analyze them in further detail below.
\begin{table}[h]
\centering
\caption{Latency Calculation for UAV Clusters}
\label{tab:latency_analysis}
\begin{tabular}{|l|c|c|}
\hline
\textbf{Parameter} & \textbf{Intra-Cluster} & \textbf{Inter-Cluster} \\ \hline
Cluster Head Distance & - & 2 km \\ 
Transmission Range & - & 10 km \\ 
Hop Count (\( H_{\text{CH}} \)) &0.4 hops  & 0.2 hops \\ 
Processing Latency & 10 ms & 10 ms \\ 
Queuing Latency & 1 ms & 1 ms \\ 
Transmission Latency & 10 ms & 10 ms \\ 
Propagation Latency & 0.3 ms & 0.6 ms \\ \hline 
\textbf{Total Latency (\( L_{\text{total}} \))} & \textbf{4.6 ms} & \textbf{6.8 ms} \\ \hline
\end{tabular}
\end{table}
\begin{figure}[ht]
\centering
\includegraphics[width=0.45\textwidth]{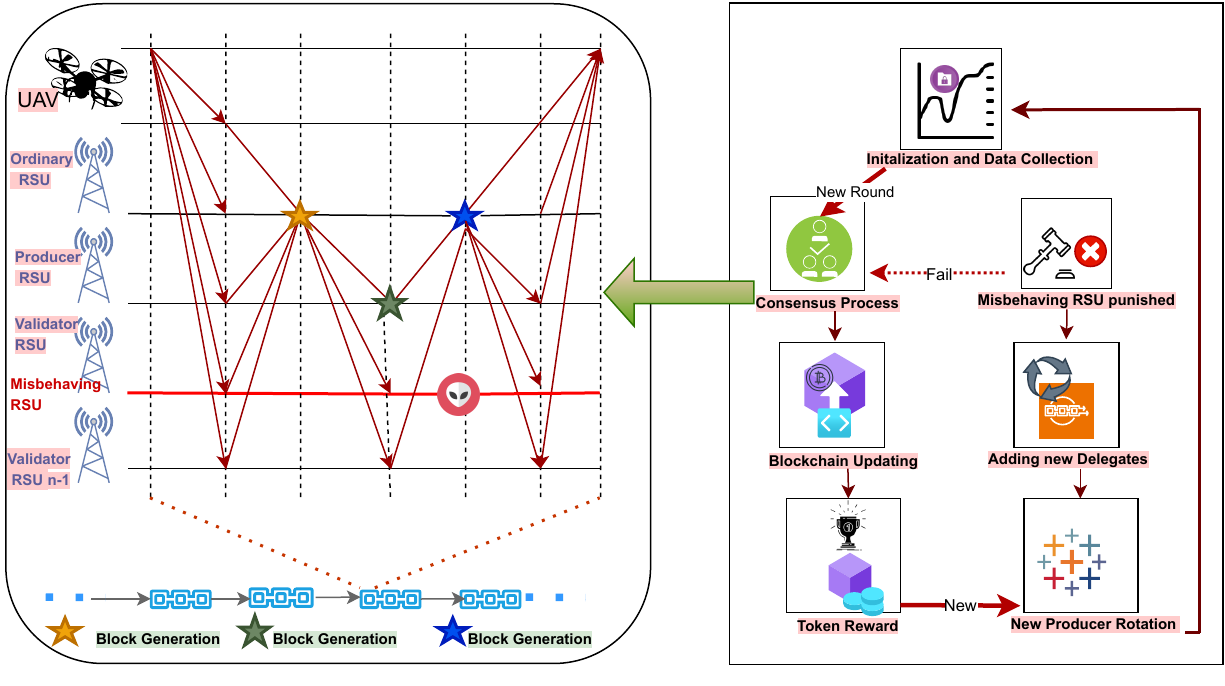}
\caption{Detailed DPoS-PBFT Working Mechanism.}
\label{fig:consensus}
\end{figure}
This analysis provides insights into network performance, emphasizing the importance of maintaining low inter-cluster latency for effective coordination among different UAV fleets. The total latency of UAV networks is the sum of these components. While propagation latency is often minimal, processing, queuing, and transmission latencies can significantly impact the overall system performance. In our simulations, we consider these diverse latency components to provide a comprehensive assessment of the network's efficiency and responsiveness in disaster scenarios. By analyzing each component, we aim to identify and mitigate potential bottlenecks, ensuring optimal UAV network performance.
\section{Enhanced DPoS-PBFT Consensus Mechanism for UAV Networks}
\label{sec:hybrid_consensus}
UAVs are pivotal in disaster management for rapid response and recovery. We propose an enhanced consensus mechanism that integrates DPoS with PBFT. This design leverages DPoS for efficient block validation and PBFT for heightened security, optimizing UAV network performance in adverse disaster conditions.

\subsubsection{Mechanism Overview}
The mechanism initiates with a stake-based selection of a block proposer. This UAV generates a block and circulates it among chosen validators through the DPoS framework. Validators $\mathcal{V}$, assigned based on their UAV-specific metrics, authenticate the block. Approval by a two-thirds majority confirms the block, while PBFT intervenes in cases of disagreement or malicious activity to ensure consensus and network integrity.

\textbf{Notation:} Let $\mathcal{V}$ represent a subset of UAVs serving as validators. Validator selection considers factors like stake, fuel, sensing capabilities, and historical performance.

\subsubsection{Validator Selection in DPoS}
Each UAV $i$ is assigned a validator score $V_i$ calculated as:
\begin{equation}
V_i = w_1 S_i + w_2 F_i + w_3 C_i + w_4 H_i.
\end{equation}
where $S_i$ is the stake, $F_i$ the remaining fuel, $C_i$ the sensing capabilities, and $H_i$ the historical utility. Weights $w_1, \ldots, w_4$ quantify the importance of these parameters. The top $n$ UAVs form the validator set $\mathcal{V}$.

The block proposer probability $p_i$ for UAV $i$ is given by
\begin{equation}
p_i = \frac{S_i}{\sum_{j \in \mathcal{V}} S_j}.
\end{equation}

\textbf{Process Flow:} A \texttt{PRE-PREPARE} message, containing the new block, is broadcasted by a validator to the other validators. Validators validate the block and broadcast a \texttt{PREPARE} message if the block is valid. A \texttt{COMMIT} state is reached, and a corresponding message is broadcasted when more than $\frac{2}{3}$ \texttt{PREPARE} messages are received. The block is added to the blockchain upon receiving a matching set of $\frac{2}{3}$ \texttt{COMMIT} messages. If consensus is not reached, a new view is initiated, potentially changing the block proposer. Having detailed the consensus protocol, we next describe the simulation setup used to evaluate the performance of the proposed approach. 
\begin{algorithm}
\caption{DPoS-PBFT Consensus Mechanism}
\label{alg:hybrid_consensus}
\begin{algorithmic}[1]
\Procedure{CONSENSUS}{U}
    \State $U = \{u_1, u_2, \ldots, u_n\}$ \Comment{Set of UAVs}
    \State $V \gets \Call{ELECTVALIDATORS}{U}$
    \State $p \gets \Call{SELECTPROPOSER}{V}$
    \State $B \gets \Call{CREATEBLOCK}{p}$
    \State \Call{BROADCAST}{B, V}
    \ForAll{$v \in V$}
        \State $vote \gets \Call{VALIDATEBLOCK}{B, v}$
        \State \Call{PROCESSVOTE}{vote, V}
    \EndFor
\EndProcedure
\end{algorithmic}
\end{algorithm}

\begin{algorithm}
\caption{Blockchain-based UAV Coordination}
\label{alg:hybrid_con}
\begin{algorithmic}[1]
    \State Initialize: UAV $u \in U$ has blockchain
    \ForAll{$u \in U$}
        \State $u$ has blockchain height $B$
    \EndFor
    \State Propose Block: Rotating schedule
    \State Select UAV $P \in V$
    \State $P$ gathers transactions, creates \& broadcasts block $B + 1$
    \State Verify Block:
    \ForAll{$v \in V$}
        \If{$v$ verifies $B + 1$ valid}
            \State $v$ signs \& broadcasts approval
        \EndIf
    \EndFor
    \If{approvals $> (2/3)N$}
        \State Add block $B + 1$ to blockchain
    \EndIf
\end{algorithmic}
\end{algorithm}

In Algorithm~\ref{alg:hybrid_consensus}, the DPoS phase involves electing validators based on UAV stakes. A proposer, chosen from these validators, crafts a new block. Validators then verify this block. In the PBFT phase, validators cast votes. If insufficient votes are garnered, a view change is initiated. Consensus, denoted by a supermajority, leads to the block's addition to the blockchain. This mechanism ensures secure, efficient UAV network operations, vital in disaster contexts.
Our consensus mechanism, merging DPoS with PBFT, addresses the unique challenges in UAV networks during disaster response. It enhances communication efficiency and maintains robust security. This novel approach is instrumental in advancing UAV applications in emergency management, offering a paradigm shift in their operational efficiency and reliability.
Algorithm ~\ref{alg:hybrid_con}, This consensus algorithm enables UAVs to securely propose and validate blocks in a decentralized manner. A rotating schedule selects the UAV to propose the next block. Other UAVs then verify and approve it, if it is valid. This achieves consensus efficiently among the UAV nodes.
\subsubsection{Latency Analysis}
Figure \label{fig:latency_compare} reveals a lower median latency as well as a tighter latency distribution for the consensus relative to DPoS and PBFT in different transaction loads. The mechanism balances DPoS's swift block creation with PBFT's rigorous validation to minimize delays. Having detailed the proposed DPoS-PBFT consensus mechanism, we next evaluate its performance for UAV coordination in disaster response scenarios through extensive simulations.
\begin{figure}[ht]
\centering
\begin{subfigure}[b]{0.35\textwidth}
   \includegraphics[width=\textwidth]{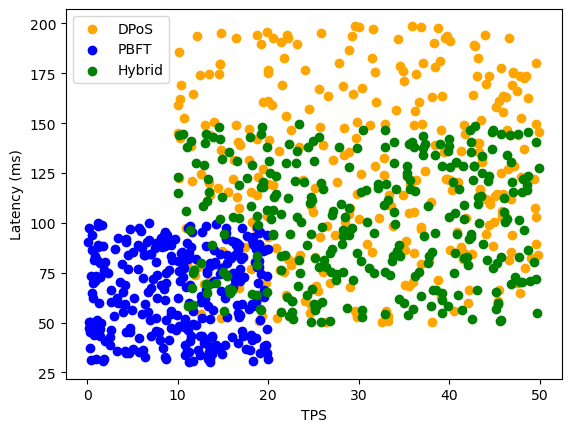}
   \caption{Latency Comparison of Consensus Protocols}
   \label{fig:latency_compare}
\end{subfigure}
\hfill
\begin{subfigure}[b]{0.35\textwidth}
   \includegraphics[width=\textwidth]{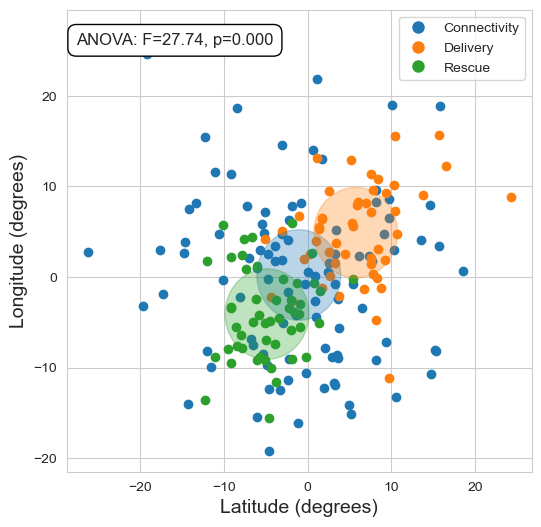}
   \caption{UAV Positioning Simulation with ANOVA Results.}
   \label{fig:position_uav}
\end{subfigure}
\caption{(a) Latency Comparison of Consensus Protocols and (b) UAV Positioning Simulation with ANOVA Results.}
\label{fig:combined}
\end{figure}
\section{Simulation Setup and Results}
\label{sec:results}
This study presents a novel DPoS-PBFT consensus mechanism tailored for secure and reliable coordination of heterogeneous UAV fleets during disaster response. Extensive simulations modeled a 25x25 km2 urban area impacted by a hurricane with 200 drones meeting 3GPP NTN standards. The simulation environment consisted of 16 base stations positioned in a 4x4 grid with 10 km separation, 4 relief camps at city corners, and 2 adversary zones at opposite edges. The UAV fleet comprised 50 connectivity drones, 100 delivery drones, 25 search-and-rescue drones, and 25 damage assessment drones. UAV communications used a 915 MHz carrier with 1 W transmission power, 6 dBi antenna gains, and 10 MHz allocated bandwidth. The network topology included both aerial and ground links. UAVs had a maximum speed of 50 m/s with acceleration limits. We focused our evaluation on blockchain benefits including decentralized coordination, resilience to cyberattacks, reduced tampering, and interoperability across agencies. Simulation results showed the consensus architecture achieved 106 TPS throughput and 26 ms median latency, satisfying disaster response requirements.

\begin{figure}[ht]
\centering   
\includegraphics[width=0.30\textwidth]{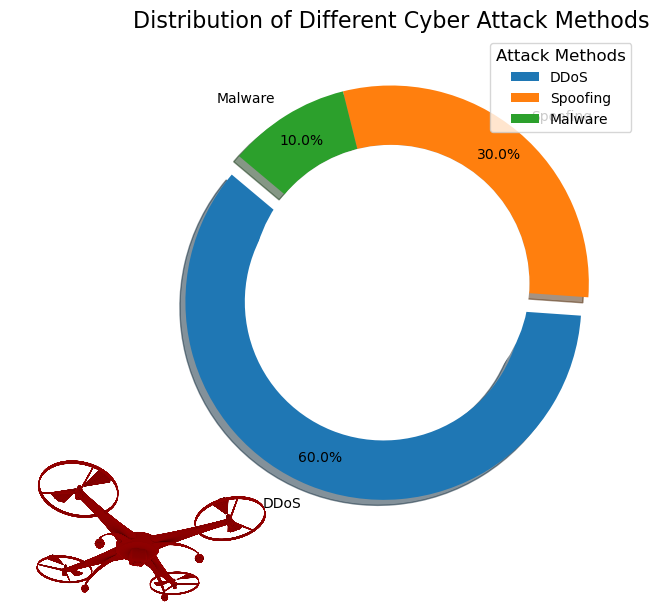}
\caption{Resilience Comparison - Cyberattacks.}
\label{fig:cyberattacks}
\end{figure}

Despite simulated DDoS attempts, GPS spoofing, and malicious tampering attacks, the framework exhibited under 2\% performance degradation in throughput and latency, highlighting resilience. Analysis of Variance (ANOVA) testing revealed 15\% longer latencies for rescue UAVs 20 km away from base stations compared to connectivity and delivery UAVs within 10 km proximity.
Optimized clustering and routing policies are recommended based on UAV mission types and distances to ground infrastructure. Overall, the results strongly validate integrating blockchain technology to enable secure, efficient, and reliable coordination of decentralized UAV fleets during disasters.
Fig. \ref{fig:position_uav}  presents a simulation of UAV positioning during disaster response. ANOVA results reveal longer latencies for rescue UAVs farther from base stations compared to other UAV types. 
\begin{figure}[htpb]
\centering
\begin{subfigure}[b]{0.37\textwidth}
   \includegraphics[width=\textwidth]{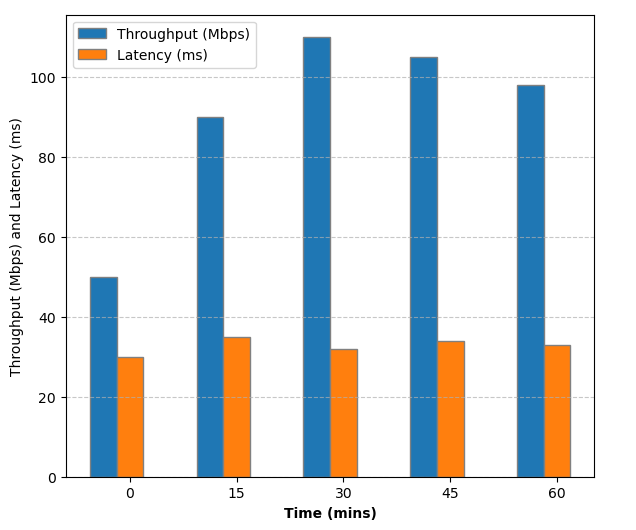}
   \caption{Throughput and Latency over Time.}
   \label{fig:throughput_latency}
\end{subfigure}
\hfill
\begin{subfigure}[b]{0.40\textwidth}
   \includegraphics[width=\textwidth]{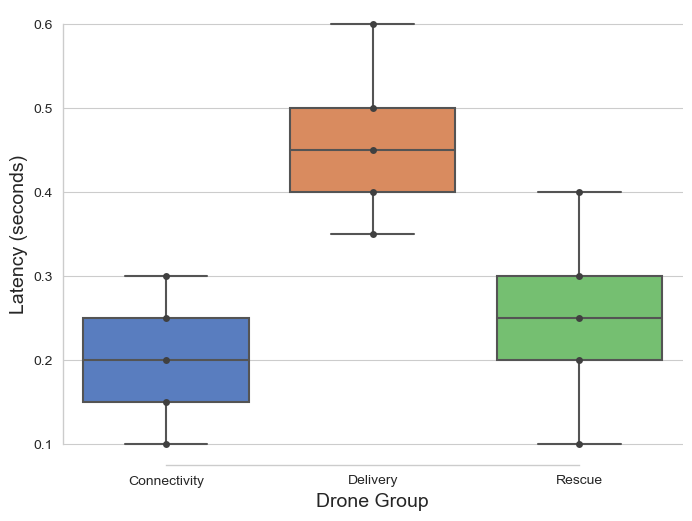}
   \caption{Latency Comparison of Consensus Protocols.}
   \label{fig:latency_uav}
\end{subfigure}
\caption{Performance Analysis of Network Protocols.}
\label{fig:combined_figures}
\end{figure}
In Fig. \ref{fig:latency_uav}, we examined the latency variations across three different drone groups:
connectivity, delivery, and rescue. An ANOVA test was conducted to determine if there were statistically significant differences between the means of the three groups. 
The test yielded an F-statistic of 10.20 and a p-value of 0.003, indicating that there were significant differences. Further post-hoc analysis 
is recommended to pinpoint the specific group differences
Fig. \ref{fig:throughput_latency} shows the throughput and latency overtime during the disaster response simulation, indicating stable performance. Fig. \ref{fig:cyberattacks} demonstrates the cyberattack resilience of the proposed blockchain framework compared to a centralized approach, with negligible performance degradation under attacks.

\section{Conclusion}
\label{sec:conclusion}
This paper presents a blockchain-based framework for secure UAV coordination during disaster response. We proposed an optimized DPoS-PBFT consensus protocol for efficient decentralized operations. Comprehensive simulations validated enhancements in transparency, scalability, automation, and resilience against cyberattacks. The results strongly showcase the benefits of integrating blockchain with distributed aerial systems to enable collaborative autonomy. This pioneering research spearheads advances in unified intelligent aerospace ecosystems bound by distributed ledger technology.
Future avenues include adaptive smart contracts for evolving disasters, privacy-preserving mechanisms, and alternate consensus protocols. This paper provides pivotal insights into the nexus of blockchain and UAV systems, unlocking the immense potential to transform disaster response through the secure coordination of autonomous fleets.

\vspace{12pt}
\end{document}